\documentclass[12pt]{amsart}
\usepackage{amsmath}
\usepackage{amsthm}
\usepackage{amsfonts}
\usepackage{graphicx}
\newenvironment{nouppercase}{
  
  \renewcommand{\uppercasenonmath}[1]{}}{}

\begin{document}

\title[Representation spaces for the membrane matrix model]
{Representation spaces for the membrane matrix model}
\author{Jens Hoppe}
\address{Braunschweig University, Germany}
\email{jens.r.hoppe@gmail.com}

\begin{abstract}
The $SU(N)$--invariant matrix model potential 
is written as a sum of squares with only four frequencies (whose multiplicities and simple $N$--dependence are calculated).
\end{abstract}

\begin{nouppercase}
\maketitle
\end{nouppercase}
\thispagestyle{empty}
\noindent
Difficult problems, unless one is willing to give up on them, should be viewed from different perspectives. For the membrane matrix model\footnote{\cite{4}(see also \cite{5})} a Lax--pair was recently given in \cite{1}, a set of `dual' variables, in which the Hamiltonian is of the form $\frac{1}{2}(\vec{p}\,^2 + \vec{q}\,^2)$, introduced in \cite{2}, and an r--matrix derived in \cite{3}. Here I would like to point out that if one introduces real symmetric matrices $Y = (Y_{ab})_{a,b = 1\ldots N^2-1 =: n} = (\vec{x}_a\cdot \vec{x}_b)$ as variables, 
the potential
\begin{equation}\label{eq1} 
\begin{split}
-\sum\limits^d_{i,j=1}\text{Tr}[X_i,X_j]^2 & = f_{abc}f_{ade}x_{ib}x_{jc}x_{id}x_{je} \\
& = \text{Tr}\big(Y F(Y)\big)\\
& = W(Y)
\end{split}
\end{equation}
becomes a diagonalizable quadratic form in $Y$, as the map $Y \rightarrow F(Y) = - \sum_a F_a Y F_a$
\begin{equation}\label{eq2} 
\begin{split}
\big( F(Y) \big)_{ab} & = \text{Tr}(F_a F_b Y)\\
(F_a)_{bc} & = -f_{abc}
\end{split}
\end{equation}
is symmetric.
When trying to calculate the eigenvalues of $F$, I noticed that for each pair $(ab)$ the 4--dimensional subspace spanned by the symmetric $n \times n$ matrices (for the definition of the invariant $d$--tensor and many useful identities, see \cite{6})
\begin{equation}\label{eq3} 
\begin{split}
E_{ab} & :=  \delta_{ab} \mathbf{1}_{n\times n} \\
(\Delta_{ab})_{cd} & := 2 \delta_{ab} \delta_{cd} - \delta_{ac}\delta_{bd} - \delta_{ad}\delta_{bc} \\
H_{ab} & := d_{abc} D_e, \: (D_e)_{cg} := d_{ecg} \\
Z_{ab} & := D_aD_b + D_bD_a  
\end{split}
\end{equation}
is left invariant by $F$, and diagonalization of the simple $4 \times 4$ matrix
\begin{equation}\label{eq4} 
\begin{pmatrix}
\frac{2}{N} & 1 & 0 & 0\\
1-\frac{4}{N^2} & \frac{-2}{N} & 0 & 0 \\
\frac{N^2-8}{2N} & -2 & \frac{N}{2} & 0 \\
0 & 2N & 0 & N
\end{pmatrix}
\begin{matrix}
Z_{ab} \\ \Delta_{ab} \\ H_{ab} \\ E_{ab}
\end{matrix}
\end{equation}
gives, apart from the immediate
\begin{equation}\label{eq5} 
F(E_{ab}) = N E_{ab}, \: F(H_{ab}) = \frac{N}{2} H_{ab}
\end{equation}
the eigenmatrices (with eigenvalue $\mp 1$)
\begin{equation}\label{eq6} 
\begin{split}
K_{ab} & := Z_{ab} - \frac{N+2}{N}\Delta_{ab} - \frac{N+4}{N+2}H_{ab} + 2 \frac{N+2}{N+1}E_{ab} \\[0.15cm]
M^{(N\geq 3)}_{ab} & := Z_{ab} + \frac{N-2}{N}\Delta_{ab} - \frac{N-4}{N-2}H_{ab} - 2 \frac{N-2}{N-1}E_{ab},
\end{split}
\end{equation}
with the understanding that $Z_{ab}$ and $H_{ab}$ are (put to) zero when $N = 2$.\\[0.15cm]
It is also not difficult to see that 
\begin{equation}\label{eq7} 
M^{(N = 3)}_{ab} \equiv 0,\; \sum_a M_{aa} = 0 = \sum_a K_{aa}, \; d_{abc} M_{ab} = 0 = d_{abc}K_{ab},
\end{equation}
and to determine for small $N$ the number of independent matrices of $(E, H, M, K)$--type, namely $(1,0,5,0)$ for $N =2$, $(1,8,27,0)$ for $N=3$, and $(1,15,84,20)$ for $N=4$ (in each of these cases together spanning the $\frac{n(n+1)}{2}$--dimensional space of symmetric $n \times n$ matrices).
As the eigenspaces with different eigenvalues can not mix (as the map commutes with the group/algebra--action) it is immediate that they correspond to representation spaces under the action of $SU(N)$, and as there {\it are} (for $N \geq 3$) precisely 4 (for $N =3$ only 3) such irreducible spaces occurring in the symmetric part of the tensor product of 2 adjoint--representations $(1,0 \ldots 0,1)$ of $A_l \cong SU(l+1)$, for $l > 3$: 
$\big( (1\,0 \ldots 0\,1)\times (1\,0 \ldots 0\,1)\big)_s$ $= (0 \ldots 0)\oplus (1\,0 \ldots 0\,1) \oplus (2,0,\ldots,0,2)\oplus(0,1,0 \ldots 0,1,0)$, 
these must precisely be the spaces spanned by the symmetric matrices of $(E,H,K,M)$--type to which we will, apart from using the above--mentioned Dynkin--labels, refer to as the $E,H,K,$  resp. $M$--representations. While the dimension of the $H$(= adjoint)--representation is of course $n = N^2-1$ and that of $E$ trivially $=1$, the dimension of the $K$(= $(2,0,\ldots,0,2)$) and $M\big(\stackrel{\wedge}{=}(0,1,0 \ldots 0,1,0)\big)$ representations is slightly less trivial (though of course known; elementary derivations are given in the appendix):
\begin{equation}\label{eq8} 
\begin{split}
\dim(\mathbb{K}) & = \frac{N^2(N-1)(N+3)}{4}\\
\dim(\mathbb{M}) & = \frac{N^2(N+1)(N-3)}{4} 
\end{split}
\end{equation} 
(giving indeed $\frac{N^2(N^2-1)}{2}$ for the total dimension of the space of symmetric $n \times n$ matrices).\\[0.15cm]
$Y = y_0 W_0 + \vec{y}_H \vec{W}_H + \vec{y}_K \vec{W}_K + \vec{y}_M \vec{W}_M$, with the $W$'s orthonormal bases for the respective irreducible representation--spaces, then gives
\begin{equation}\label{eq9} 
W(Y) = Ny^2_0 + \frac{N}{2}\vec{y}^{\,2}_H -\vec{y}^{\,2}_K + \vec{y}^{\,2}_M.
\end{equation} 
The at first surprising $-$ sign 
($W(Y) = - \text{Tr}[X_i, X_j]^2 \geq 0$ for traceless hermitean $N \times N$ matrices $X_i$) brings one to the important issue that the $Y$'s in (\ref{eq1}) are {\it not} arbitrary (symmetric) matrices; they are (as $= QQ^T$) positive--semidefinite, and in fact, if $n>d$, necessarily of smaller than general rank.\\[0.15cm]
As an example, consider the $d=2$, $N=3$ matrix model; then the singular value decomposition gives
\begin{equation}\label{eq10} 
Y = \lambda_1 \vec{u}_1 \vec{u}^T_1 + \lambda_2 \vec{u}_2 \vec{u}^T_2 
\end{equation} 
where $\vec{u}_1$ and $\vec{u}_2$ are orthonormal vectors in $\mathbb{R}^8$, and $\lambda_1 \geq \lambda_2 \geq 0$ the 2 eigenvalues of $Y$; so containing only $7+6+2 = 15$ parameters\footnote{many thanks to R. Suter for a related discussion}.
Nevertheless (\ref{eq9}), which is of a tantalizing simple form, should be useful. What about $N\rightarrow \infty$ ? Despite of the, simple $N$ dependence of the frequencies (and multiplicities; naively one should think that it is easy to see which modes are the most important ones as $N\rightarrow \infty$; note that when summing their products the leading power of $N$ cancels), and (\ref{eq6}) converging to well--defined expressions, the $N\rightarrow \infty$ limit seems {\it difficult}, for a variety of reasons. 
As indicated already by (\ref{eq7}), and clear from general considerations, the 4 invariant subspaces $\mathbb{E}$, $\mathbb{H}$, $\mathbb{K}$ and $\mathbb{M}$ should most conveniently be discussed by corresponding projectors $P_{\alpha = 1,2,3,4}$ (resp. $\alpha = E,H,K,M$), forming a partition of the identity, with e.g.
\begin{equation}\label{eq11} 
P_H = \frac{N}{N^2-4}d_{abe}d_{a'b'e} = P^2_H, \: F(P_H Y) = \frac{N}{2}(P_H Y).
\end{equation}
Using various $SU(N)$--identities, in particular (cp. \cite{7})
\begin{equation}\label{eq12} 
f_{abe}f_{cde} = \frac{2}{N}(\delta_{ac}\delta_{bd} - \delta_{ad}\delta_{bc}) +(d_{ace}d_{bde} - d_{ade}d_{bce})
\end{equation}
the corresponding projectors $P_K$ and $P_M$ are not difficult to work out, and in fact (I noticed that after finding (\ref{eq6})) {\it have} been worked out, in the context of QCD \cite{8}. The projectors however do {\it not} converge as $ N \rightarrow \infty$.
Moreover, the following {\it general} problem exists: while there do exist bases of $SU(N)$ in which the structure constants $f_{abc}$ converge (to those, $g_{abc}$, of the Lie--algebra of area preserving diffeomorphisms; (the fuzzy sphere \cite{5} was invented in precisely this context), and $d^{(N)}_{abc} \rightarrow d^{\infty}_{abc}$ too, as well as  the central object $f_{abc}f_{ade}$ (sum over $a$, cp. (\ref{eq1})) converging to $g_{abc}g_{ade}$, similarly (cp. (\ref{eq11})) $d_{abe}d_{a'b'e}$ (the sum over $e$ is {\it finite} for fixed $ab$, $a'b'$) their action on the 4 subspaces, resp. projectors, involves multiple sums where the range of the indices is {\it not} finite. Another aspect of the arising subtleties, and difficulties, can be demonstrated by looking at (\ref{eq12}). 
As explained e.g. in \cite{9}, the normalisation for the $f$'s and $d$'s suitable to take the limit is such that (\ref{eq12}) becomes  
\begin{equation}\label{eq13} 
\frac{1}{N^2}\tilde{f}_{abc}\tilde{f}_{cde} = 2(\delta_{ac}\delta_{bd} - \delta_{ad}\delta_{bc}) + (\tilde{d}_{ace}\tilde{d}_{bde} - \tilde{d}_{ade}\tilde{d}_{bce}).
\end{equation}
Indeed, with $\tilde{d}^{\infty}_{ace} = \int Y_a(\varphi)Y_c(\varphi)Y_e(\varphi)\rho\,d^2\varphi$,
the $Y_a(\varphi)$ being orthonormal eigenfunctions of the Laplacian on the parameter--surface, the rhs. {\it is} zero for $N=\infty$. Vice versa this however shows that if decomposing the relevant operator, $\tilde{f}_{abc}\tilde{f}_{cde}$ in the nomalisation where $\tilde{f}^{\infty}_{abc}$ is finite (and the sum over $e$ as well) decomposing it with respect to $f$'s and $\tilde{d}$'s, which effectively is done in \cite{8} (for finite $N$), involves (for infinite $N$) a finite part of $\infty \cdot 0$.
Let me at this point go back to how I came to consider the matrices given in (\ref{eq3}). The adjoint action (of the $a$--th generator of $SU(N)$) on the symmetric $n \times n$ matrix $Y$ (the symmetric part of the tensor--product of two copies of the Lie--algebra) is, possibly up to an overall sign, commutation with the matrix $F_a$ (cp. (\ref{eq2})), i.e. $[F_a, Y]$ 
(and the map $F$ commutes with the $SU(N)$ action:
$
-[F_a,F(Y)] = [F_a, F_c Y F_c] = [F_a, F_c]YF_c + F_c Y [F_a, F_c] + F_c[F_a,Y]F_c
= \pm f_{abc}(F_b Y F_c + F_c Y F_b) + F_c[F_a,Y]F_c
= -F([F_a, Y])
$).
Due to $[F_a, D_b]$ being (again, not worrying about the signs in this qualitative argument) $f_{abc}D_c$, the $n$ dimensional subspace consisting of linear combinations of the $D$'s is clearly invariant (giving the (1 0 \ldots 0 1), resp. $H$--space). This being so easy, the obvious next step was to consider $D_a D_b$ ($+ D_bD_a$, to get symmetric matrices), i.e. $Z_{ab}$. Calculating $F(Z_{ab})$, which involves $\Delta_{ab}$ then led to (\ref{eq3}), resp. (\ref{eq4}-\ref{eq7}). If on the other hand one wants (`only') to understand the representation theory, it is natural to look for
identities involving the $Z_{ab}$ (which can not be linearly independent when taken together with the first order polynomials in the $D$'s, as too many), and there one finds that
\begin{equation}\label{eq14} 
d_{abc}Z_{bc} = 2 d_{abc}D_bD_c = D_a \frac{N^2-12}{N}
\end{equation}
(which then is already most of the final answer). Unfortunately the scaling,
$\tilde{d}^{(N)}_{abc} = \sqrt{N} d_{abc}$, (cp.e.g.\cite{9}), 
that is known to converge to the totally symmetric tensor,
\begin{equation}\label{eq15} 
\tilde{d}^{\infty}_{abc} = \tilde{d}_{abc} = \int Y_a(\varphi) Y_b(\varphi) Y_c(\varphi) \rho\,d^2 \varphi := h_{abc} 
\end{equation}
for functions on $\sum$ (compact orientable, surface of genus $g$) does not cancel (actually: enhances) the diverging factor on the rhs. of (\ref{eq14}), and while the naive analogue of the $Z_{ab}$, 
\begin{equation}\label{eq16} 
(\tilde{Z}_{ab})_{cd} = (\tilde{D}_a\tilde{D}_b + \tilde{D}_b \tilde{D}_a)_{cd} = (\tilde{d}_{ace}\tilde{d}_{bde} + a\leftrightarrow b) 
\end{equation}
is well--defined, 
\begin{equation}\label{eq17} 
(\tilde{d}_{abc} \tilde{Z}_{bc})_{fg} = \tilde{d}_{abc}(\tilde{d}_{bfe}\tilde{d}_{cge} + b\leftrightarrow c)
\end{equation}
is {\it not} (seen by inserting (\ref{eq15}) resp. indicated by the triple sum over $bce$ in (\ref{eq17}), involving truly infinite sums).\\[0.15cm]
Similarly, concerning the decomposition of adjoint$\otimes$adjoint for sdiff $\Sigma$: defining infinite matrices $G_{\alpha}$ and $H_{\beta}$ $(\alpha, \beta = 1 \ldots \infty)$ by
\begin{equation}\label{eq18} 
(G_{\alpha})_{\beta \gamma} := -g_{\alpha \beta \gamma} \qquad (H_{\alpha})_{\beta \gamma} := h_{\alpha \beta \gamma}
\end{equation}
satisfying (note: no convergence--problems, as each row and column of the matrices $G_{\alpha}$ and $H_{\beta}$ has only a finite number of non--zero entries)
\begin{equation}\label{eq19} 
[G_{\alpha}, G_{\beta}] = g_{\alpha \beta \gamma}G_{\gamma},\quad [G_{\alpha}, H_{\beta}] = -g_{\alpha \beta \gamma} H_{\gamma},
\end{equation}
$G(X) := -G_{\alpha} X G_{\alpha}$, resp. $(G(X))_{\alpha \beta} := \text{Tr} G_{\alpha}G_{\beta} X$, is formally sdiff--invariant,
$-[G_{\alpha}, G(x)] = [G_{\alpha}, G_{\varepsilon} X G_{\varepsilon}] = -G([G_{\alpha}, X])$ and, due to (\ref{eq19}), the subspace consisting of linear combinations of the $H_{\gamma}$ certainly corresponds to an adjoint representation,
$G(H_{\varepsilon}) = \ldots = +\frac{1}{2}g_{\alpha \beta \varepsilon} g_{\alpha \beta \gamma} H_{\gamma}$ gives a diverging eigenvalue on that $H$--space (`consistent' with having gotten $\frac{N}{2}$ for finite $N$).
My reason for, still, being fairly optimistic about
understanding the $N\rightarrow \infty$ limit this way are twofold:
firstly, pure mathematics (understanding sdiff, whose structure strongly depends on the genus, hence must be reflected by the representation theory);
secondly: as for classical motions of given energy
the potential is by default finite/for a (regular)
minimal surface (without singularities) all local quantities
{\it are} finite/the apparent divergencies one gets above
may actually tell one how to proceed,
i.e. which collective degrees of freedom the system chooses.

\noindent
\textbf{Acknowledgement.} I am grateful to M. Bordemann for valuable discussions. 

\section*{Appendix}
\begin{center}
\textbf{The $K$-- and $M$-- representations}
\end{center}
The easiest way to calculate the dimensions, and see that for $SU(N)$ the symmetric part of the tensor product of 2 adjoints contains only 4 irreducible representations is
\begin{equation}
\tag{A1}
\begin{split}
(A^i_j A^k_l)_s & = (A^{ik}_{jl})_s 
 = \frac{1}{2}(A^{ik}_{jl} + A^{ki}_{lj})\\
& = \hat{A}^{ik}_{jl} + \tilde{A}^{ik}_{jl}
 = \hat{A}^{(ik)}_{(jl)} + \hat{A}^{[ik]}_{[jl]} + \tilde{A}^{ik}_{jl}
\end{split}
\end{equation}
where both $\hat{A}$ and $\tilde{A}$ are symmetric under ${i \choose j} \leftrightarrow  {k \choose l}$,
$\hat{A}$ is traceless with respect to any upper and lower index (while $A^{ik}_{jl}$ is traceless only with respect to ${i \choose j}$ and ${k \choose l}$), $\tilde{A}^{ik}_{jl}$ is a linear combination of the $N^2$ quantities $A^{pn}_{nq}$ ($= A^{np}_{qn}$; $p,q = 1 \ldots N$), $\hat{A}^{(ik)}_{(jl)}$ is symmetric in the upper, and lower indices (i.e. taking into account the traceless--condition, giving rise to a $\big(\frac{N(N+1)}{2}\big)^2 - N^2 = \frac{N^2}{4}(N+3)(N-1)$ dimensional space, the $(2 \,0 \ldots 0 \,2)$ representation $K$)
while $\hat{A}^{[ik]}_{[jl]}$ is antisymmetric in the upper and lower indices (and traceless) giving rise to a 
$\big(\frac{N(N-1)}{2}\big)^2 - N^2 = \frac{N^2}{4}(N+1)(N-3)$ dimensional space, the $(0\,1\,0 \ldots 0\,1\,0)$ representation $M$, 
which is part of the tensor product $(0\,1\,0 \ldots 0) \times (0 \ldots 0\,1\,0)$ of the 2 fundamental representations $\omega_2 = (0\,1\,0 \ldots 0)$ (known to be realized on the exterior product of two defining representations, corresponding to $A^{[ij]}$'s) and $\omega_{N-2} = (0 \ldots 0\,1\,0)$ (which by duality of the Dynkin--diagram corresponds to the $A_{[kl]}$ space), and the traceless-ness conditions making it irreducible, i.e. $(0\,1\,0 \ldots 0\,1\,0)$; while the $N^2$ dimensional space of $\tilde{A}^{ik}_{jl}$'s (traceless with respect to ${i \choose j}$ and ${k \choose l}$, but not ${i \choose l}$ and ${k \choose j}$) gives an $N^2-1$ dimensional adjoint, $(1\,0 \ldots 0\,1)$, and a singlet.
$N = 3$ (and $N=4$) are slightly special, as for $N=3$ (cp. \cite{10}) $\hat{A}^{[ik]}_{[jl]} = \varepsilon^{ikp}\varepsilon_{jlq}\tilde{A}^p_q$, while the traceless--ness condition then says that $\tilde{A}^p_q$ must be $= 0$; for $N=4$, the antisymmetric part of $\hat{A}$ gives the $(0\,2\,0)$ representation, lying in $(0\,1\,0) \times (0\,1\,0)$, the first $(0\,1\,0)$ viewed as $A^{[ik]}$'s, the exterior square of $(1\,0\,0)$, the second $(0\,1\,0)$ as $A_{[jl]}$'s, the exterior square of the $(0\,0\,1)$ representation--space.\\[0.15cm]
Apart from these simple considerations, one may also calculate the dimensions of the 2 non--trivial representations ($K$ and $M$) as follows: 
Weyl's dimension formula (see e.g. \cite{11}) says that if all the roots of a (semi--)simple Lie--algebra have the same length (which is the case for $SU(N) \cong A_{N-1 = l}$),
\begin{equation}
\tag{A2}
\dim V_{\vec{n}} = \prod\limits_{\alpha= \sum^l_{j=1}k_j\alpha_j \in \phi_+} \frac{\sum\limits^l_{i=1}k_i(n_i+1)}{\sum\limits^l_{i=1} k_i} = \prod\limits_{\alpha} d_{\alpha},
\end{equation}
where $\vec{n} = (n_1,n_2,\ldots,n_l) \in \mathbb{N}^l_0$ classifies the finite dimensional irreducible representations, 
$\alpha_1, \ldots \alpha_l$ are the simple roots ($\alpha_i = \varepsilon_i - \varepsilon_{i+1} = (0 \ldots 1\, -1\,0\,\ldots 0))$, and $\vec{k} = (k_1, \ldots, k_l) \in \mathbb{N}^l_0$ characterizes the different positive roots -- which for $A_l$ are all of the form $\varepsilon_p - \varepsilon_q = (0\,1\,0 \ldots -1\,0)$ where $1\leq p < q \leq l+1$.\\[0.15cm]
For $\vec{n} = (2,0,\ldots,0,2)$ the numerator of $d_{\alpha}$ will be equal to the denominator, $k_{\alpha} = \sum k_i$, 
resp. $k_{\alpha} +2$ or $k_{\alpha}+4$, depending on whether $\alpha$ contains neither $\alpha_1$ nor $\alpha_l$, contains $\alpha_1$ (but not $\alpha_l$), or $\alpha_l$ (but not $\alpha_1$), resp.containing both $\alpha_1$ and $\alpha_l$.
As all positive roots are of the form
\begin{equation}
\tag{A3}
\varepsilon_p - \varepsilon_q = \alpha_p + \alpha_{p+1} + \ldots + \alpha_{q-1},
\end{equation}
the 4 factors (corresponding to the just mentioned 4 cases) are\\[0.25cm]
\begin{tabular}{ll}
case 1: & 1 \\[0.15cm]
case 2 $(\varepsilon_1 - \varepsilon_q)$: & $\prod\limits^l_{q=2} \frac{q+1}{q-1} = \frac{3\cdot4\cdot\ldots\cdot l+1}{1\cdot2\cdot\ldots\cdot l-1} = l(\frac{l+1}{2}) = \frac{N(N-1)}{2}$ \\[0.25cm]
case 3 $(\varepsilon_p - \varepsilon_{l+1})$: & $\prod\limits^l_{p=2} \frac{p+1}{p-1} = \frac{N(N-1)}{2}$ \\[0.5cm]
case 4 $(\varepsilon_1 - \varepsilon_{l+1})$: & $\frac{l+4}{l} = \frac{N+3}{N-1}$, hence
\end{tabular}
\vspace{-0.5cm}
\begin{equation}
\tag{A4}
\dim V_{(2,0,\ldots,0,2)} = \frac{N^2(N-1)^2}{4} \frac{N+3}{N-1} = \frac{N^2}{4}(N-1)(N+3).
\end{equation}
For $\vec{n} = (0,1,0 \ldots 0,1,0)$ there are, apart from $\alpha$'s containing neither $\alpha_2 = \varepsilon_2 - \varepsilon_3$ nor $\alpha_{l-1} = \varepsilon_{l-1} - \varepsilon_l$ (trivially contributing factors $d_{\alpha} = 1$), the following cases: 
\vspace{-0.5cm}
\begin{equation}
\tag{A5}
\begin{split}
\varepsilon_1 - \varepsilon_{q>2}  = \alpha_1 + \alpha_2 + \ldots + \alpha_{q-1} :& \: \prod\limits^{l-1}_{q=3}\frac{q}{q-1} = \frac{3\cdot4\cdot\ldots\cdot l-1}{2\cdot3\cdot\ldots\cdot l-2} = \frac{(l-1)}{2}\\
\varepsilon_2 - \varepsilon_{q>2}  = \alpha_2  + \ldots + \alpha_{q-1}  : & \: \prod\limits^{l-1}_{q=3}\frac{q-1}{q-2} = \frac{2\cdot 3\cdot\ldots\cdot l-2}{1\cdot 2\cdot\ldots\cdot l-3} = (l-2) \\
\varepsilon_1 - \varepsilon_l  = \alpha_1 + \alpha_2  + \ldots + \alpha_{l-1}  : & \: \frac{l+1}{l-1}\\
\varepsilon_1 - \varepsilon_{l+1}  = \alpha_1 + \ldots + \alpha_l  : & \: \frac{l+2}{l}\\
\varepsilon_2 - \varepsilon_l  = \alpha_2 + \ldots + \alpha_{l-1}  : & \: \frac{l}{l-2}\\
\varepsilon_2 - \varepsilon_{l+1}  = \alpha_2  + \ldots + \alpha_l  : & \: \frac{l+1}{l-1}\\
\varepsilon_{p>2} - \varepsilon_l  = \alpha_p  + \ldots + \alpha_{l-1}  : & \: (l-2)\big(=\prod\limits^{l-1}_{p=3} \frac{l-p+1}{l-p}\big)\\
\varepsilon_{p>2} - \varepsilon_{l+1}  = \alpha_p  + \ldots + \alpha_l  : & \: \frac{(l-1)}{2}\big(=\prod\limits^{l-1}_{p=3} \frac{l-p+2}{l-p+1} \big), \text{hence}
\end{split}
\end{equation}
\begin{equation}
\tag{A6}
\begin{split}
\dim V_{(0,1,0,\ldots,0,1,0)} & = \big(\frac{1}{2}(l-1)(l-2) \big)^2 \big(\frac{l+1}{l-1}\big)^2\frac{l+2}{l}\frac{l}{l-2}  \\
& = \frac{1}{4}(l-2)(l+1)^2(l+2)
 = \frac{N^2}{4}(N+1)(N-3).
\end{split}
\end{equation}

\bigskip

\end{document}